\documentclass[twocolumn,floatfix]{revtex4}
\pdfoutput=1
\usepackage{graphicx}
\usepackage{dcolumn}
\usepackage{longtable}
\usepackage{amsmath}
\usepackage{amssymb}
\usepackage{float}
\usepackage{adjustbox}
\usepackage{epstopdf}
\usepackage[version=3]{mhchem}
\usepackage{braket}
\usepackage{latexsym}
\usepackage{float}
\usepackage{subfigure}
\usepackage{young}
\usepackage[dvipsnames]{xcolor}
\usepackage[toc,page]{appendix}
\usepackage{bbm}

\usepackage{tikz}
\usetikzlibrary{positioning,shapes,fit,arrows}
\definecolor{myblue}{RGB}{56,94,141}

\begin{document}

\title
{Connecting the proxy-SU(3) symmetry to the shell model}

\author
{Dennis Bonatsos$^1$, Andriana Martinou$^1$, I. E. Assimakis$^1$, S.K. Peroulis$^1$,   S. Sarantopoulou$^1$, and N. Minkov$^2$}

\affiliation
{$^1$Institute of Nuclear and Particle Physics, National Centre for Scientific Research 
``Demokritos'', GR-15310 Aghia Paraskevi, Attiki, Greece}

\affiliation
{$^2$Institute of Nuclear Research and Nuclear Energy, Bulgarian Academy of Sciences, 72 Tzarigrad Road, 1784 Sofia, Bulgaria}

\begin{abstract}

Proxy-SU(3) symmetry is an approximation scheme extending the Elliott SU(3) algebra of the sd shell to heavier shells. When introduced in 2017, the approximation had been justified by calculations carried out within the Nilsson model. Recently our group managed to map the cartesian basis of the Elliott SU(3) model onto the spherical shell model basis, proving that the proxy-SU(3) approximation corresponds to the replacement of the intruder orbitals by their de Shalit-Goldhaber partners, paving the way for using the proxy-SU(3) approximation in shell model calculations.  The connection between the proxy-SU(3) scheme and the spherical shell model has also been worked out in the original framework of the Nilsson model, with identical results.   

\end{abstract}

\maketitle

\section{SU(3) symmetry in nuclear structure}

The SU(3) symmetry has been used in nuclear structure for a  long time \cite{Kota}. 

In 1949 the shell model has been introduced \cite{Mayer1,Mayer2,Jensen,MJ}, which is based on a three-dimensional (3D) isotropic harmonic oscillator (HO), with a spin-orbit term added to it. The 3D isotropic HO is known to possess shells labeled by the number of excitation quanta $n$, characterized by the unitary symmetries U((n+1)(n+2)/2), having SU(3) subalgebras \cite{Wybourne,Smirnov,IacLie,BK}. 

The shell model was considered adequate for describing near-spherical nuclei, thus in 1952 the collective model of Bohr and Mottelson has been introduced \cite{Bohr,BM}, in which departure from the spherical shape and from axial symmetry are described by the collective variables $\beta$ and $\gamma$ respectively. 

A modified version  of the shell model, allowing for axial nuclear deformation to be included, has been introduced by Nilsson in 1955 \cite{Nilsson1,NR}, based on a 3D anisotropic HO with cylindrical symmetry \cite{Takahashi,Asherova,RD,ND,PVI,Lenis}. 

In 1958 Elliott proved that deformation within the nuclear sd shell with U(6) symmetry can be described in terms of its SU(3) subalgebra \cite{Elliott1,Elliott2,Elliott3,Elliott4,Harvey}. 

Early attempts since 1972 of extending the SU(3) symmetry to heavy nuclei \cite{Raychev25,Afanasev,Raychev16,RR27} evolved into the Vector Boson Model \cite{Minkov1,Minkov2,Minkov3}, while at the same time the group theoretical structure of the Bohr--Mottelson model, having an overall U(5) symmetry possessing an O(5) subalgebra, has been understood \cite{AR}. 

Beyond the sd nuclear shell the SU(3) symmetry of the 3D isotropic HO is known to be broken by the spin-orbit force, which within each HO shell pushes the orbitals possessing the highest angular momentum $j$ to the shell below. As a consequence, each shell consists by the orbitals left back after this removal, called the normal parity orbitals, plus the orbitals invading from the shell above, having the opposite parity and called the intruder orbitals. 

A major step forward has been taken in 1973, with the introduction of the pseudo-SU(3) symmetry \cite{Adler,Shimizu,pseudo1,pseudo2,DW1,DW2,Harwood}, applicable in heavy nuclei. Within the pseudo-SU(3) framerwork, the SU(3) symmetry of the 3D isotropic HO is recovered by mapping the incomplete set of normal parity orbitals left in a shell onto the complete set of orbitals of the shell below. This becomes possible by assigning to each orbital a pseudo-orbital angular momentum and a pseudo-spin, while the total angular momentum remains intact. It has been realized later that this mapping is equivalent to a unitary transformation \cite{AnnArbor,Quesne,Hess}, and the relativistic mean field origins of the  pseudospin symmetry have been understood \cite{Ginocchio1,Ginocchio2}. Within the pseudo-SU(3) scheme each shell consists of a  normal parity part, which possesses a U(n) symmetry and a SU(3) subalgebra, and an intruder part which does not possess any SU(3) structure and has to be treated separately by shell model techniques \cite{DW1,DW2}. 

In 1974 it was realized that the nuclear quadrupole degree of freedom can be described in terms of a SU(6) algebra \cite{Jolos}. 

Next year the Interacting Boson Model \cite{AI,IA,IVI,FVI} has been introduced, which also has an overall U(6) symmetry built by $s$-bosons of zero angular momentum and $d$-bosons of angular momentum two, possessing three limiting symmetries, U(5) for vibrational nuclei, which is equivalent to the Bohr-Mottelson collective model, O(6) for $\gamma$-unstable nuclei, and SU(3) for deformed nuclei. 

A SU(3) limiting symmetry, appropriate for deformed nuclei, has also been obtained in 1980 within the symplectic model \cite{Rosensteel,RW}, which uses fermion pairs, as well as in 1982 within the Interacting Vector Boson Model \cite{IVBM1,IVBM2}, in which two vector bosons of angular momentum one are used, and in 1987 within the Fermion Dynamical Symmetry Model \cite{FDSM}, in which the total angular momentum of the nucleons is assumed to be split into active and inactive parts instead of orbital angular momentum and spin parts. 
  
In 2017 the proxy-SU(3) symmetry has been introduced \cite{proxy1,proxy2,proxy3}, in which the intruder orbitals in each shell (except the one with the highest projection of the total angular momentum) are replaced by the orbitals which have deserted this shell by sinking into the shell below. As a result of this replacement, each shell regains the relevant U(n) symmetry having a SU(3) subalgebra, with only one orbital (which can accommodate two particles) remaining estranged. However, this orbital is the one lying highest in energy within the shell, thus it should be empty for most of the nuclei living in this shell. Therefore it is expected that its influence on the structure of most nuclei living in the shell should be minimal.

The proxy-SU(3) scheme has been initially justified as a good approximation through calculations \cite{proxy1} carried out within the Nilsson model \cite{Nilsson1,NR}. In the present work we are going to discuss its justification through its connection to the shell model. However, before doing so, it is instructive to discuss the nature of nucleon pairs related to the development of nuclear deformation.     
  
\section{Nucleon pairs favoring deformation} 

As early as 1953 it has been observed by deShalit and Goldhaber \cite{deShalit} in their studies of $\beta$ transition probabilities that within the proton--neutron pairs of orbitals (1p3/2, 1d5/2), (1d5/2, 1f7/2), (1f7/2,   1g9/2), (1g9/2, 1h11/2), (1h11/2, 1i13/2) the nucleons of one kind (protons, for example) have a stabilizing effect on pairs of nucleons of the other kind (neutrons in the example), thus favoring the development of nuclear deformation. In the standard shell model notation $\vert n l j m_j\rangle$, in which states are labeled by the number of oscillator quanta $n$, the orbital angular momentum $l$, the total angular momentum $j$, and its $z$-projection $m_j$, the orbitals forming a pair differ by   $\vert \Delta n \Delta l \Delta j \Delta m_j\rangle = \vert 0 1 1 0\rangle$. 

A major step forward in our understanding of effective interactions and coupling schemes in nuclei has been taken in 1962 by Talmi \cite{Talmi62} through the introduction of seniority \cite{Talmi62,Talmi71,Talmi73,Talmi93}, representing the number of nucleon pairs coupled to non-zero angular momentum, which explained the linear dependence of neutron separation energies on the mass number within series of isotopes. 

\begin{table}[htb]
\centering
\caption{Pairs of orbitals playing a leading role in the development of deformation in different mass regions of the nuclear chart according to Federman and Pittel \cite{FP1,FP2,FP3}. The pairs on the left part of the table contribute in the beginning of the relevant shell, while the pairs on the right become important further within the shell. Adapted from Ref. \cite{EPJASM}.} \label{FPpairs}
\begin{tabular}{ r r r r r   }
\hline\noalign{\smallskip}
 &  protons  & neutrons &      protons  &     neutrons  \\ 
\noalign{\smallskip}\hline\noalign{\smallskip}
light       & 1d$^{5/2}$     & 1d$^{3/2}$    &    1d$^{5/2}$      &  1f$^{7/2}$        \\
intermediate& 1g$^{9/2}$     & 1g$^{7/2}$    &    1g$^{9/2}$      &  1h$^{11/2}$       \\
rare earths & 1h$^{11/2}$    & 1h$^{9/2}$    &    1h$^{11/2}$     &  1i$^{13/2}$       \\
actinides   & 1i$^{13/2}$    & 1i$^{11/2}$   &    1i$^{13/2}$     &  1j$^{15/2}$       \\

\noalign{\smallskip}\hline
\end{tabular}
\end{table}

In 1977 Federman and Pittel \cite{FP1,FP2,FP3} realized that when adding valence protons and valence neutrons to a nucleus, the proton--neutron pairs (1d5/2, 1d3/2), (1g9/2, 1g7/2), (1h11/2, 1h9/2), and (1i13/2, 1i11/2) are responsible for the onset of deformation, while deformation is then established by the proton--neutron pairs (1d5/2, 1f7/2), (1g9/2, 1h11/2), (1h11/2, 1i13/2), and (1i13/2, 1j15/2), shown in Table \ref{FPpairs}. 
In the shell model notation these sets correspond to $\vert \Delta n \Delta l \Delta j \Delta m_j\rangle = \vert 0 0 1 0\rangle$ and $\vert 0 1 1 0\rangle$ respectively, the latter set coinciding with the de 
Shalit--Goldhaber pairs. 

The decisive role played by proton-neutron pairs has been demonstrated in 1985 through the introduction of the $N_p N_n$ scheme \cite{CastenPRL,Casten} and the $P$-factor, $P= N_p N_n / (N_p+N_n)$ \cite{Haustein,Castenbook}, by showing the systematic dependence of several observables on the competition between the quadrupole deformation, ``measured'' by the quadrupole-quadrupople interaction through $N_pN_n$, and the pairing interaction, ``measured'' through $N_p+N_n$, where $N_p$ ($N_n$) is the number of valence protons (neutrons). 

In 1995 the quasi-SU(3) symmetry \cite{Zuker1,Zuker2} has been introduced, based on the proton--neutron pairs (1g9/2, 2d5/2), (1h11/2, 2f7/2), (1i13/2, 2g9/2), expressed as $\vert \Delta n \Delta l \Delta j \Delta m_j\rangle = \vert 1 2 2 0\rangle$ in the shell model notation, which lead to enhanced quadrupole collectivity \cite{Kaneko}. 

Following detailed studies of double differences of binding energies \cite{Cakirli94,Cakirli96,Brenner,Stoitsov,Cakirli102}, in 2010 it has been realized \cite{Burcu} that proton-neutron pairs differing by $\Delta K[ \Delta N \Delta n_z \Delta \Lambda]=0[110]$ in the Nilsson notation \cite{Nilsson1,NR} $K [N n_z \Lambda]$, where $N$ is  the total number of oscillator quanta, $n_z$ is the number of quanta along the $z$-axis, and $\Lambda$, $K$ are the projections along the $z$-axis of the orbital angular momentum and the total angular momentum respectively, play a major role in the development of nuclear deformation, due to their large spatial overlaps \cite{Sofia}. These pairs correspond to the replacements made within the proxy-SU(3) scheme \cite{proxy1,proxy2,proxy3}. No relation to the pairs mentioned in the previous paragraphs had been realized at that time. 

\begin{table*}
\caption{The transformation matrix $R\cdot C$ for $N=2$. The shell model orbitals appear in the first line, while the Elliott orbitals appear in the first column. These orbitals are used in the harmonic oscillator shell 8-20 ($sd$ shell), or in the proxy-SU(3) shell 14-26 after the replacement of the intruder orbitals with their de Shalit--Goldhaber partners. Adapted from Ref. \cite{EPJASM}.}\label{N2}
\begin{tabular}{lll}
$\begin{array}{ccccccccccccc}
\hline\noalign{\smallskip}\\
\ket{n,l,j,m_j} & \ket{2s^{1/2}_{-1/2}} & \ket{2s^{1/2}_{1/2}} & \ket{1d^{3/2}_{-3/2}} & \ket{1d^{3/2}_{-1/2}} &  \ket{1d^{3/2}_{1/2}} & \ket{1d^{3/2}_{3/2}} & \ket{1d^{5/2}_{-5/2}}& \ket{1d^{5/2}_{-3/2}} & 
\ket{1d^{5/2}_{-1/2}} & \ket{1d^{5/2}_{1/2}} & \ket{1d^{5/2}_{3/2}} & \ket{1d^{5/2}_{5/2}}\\
\ket{n_z,n_x,n_y,m_s}&&&&&\\ \noalign{\smallskip}\hline\noalign{\smallskip}
\ket{0,0,2,-\frac{1}{2}}& -\frac{1}{\sqrt{3}} & 0 & 0 & -\frac{1}{\sqrt{15}} & 0 & -\frac{1}{\sqrt{5}} & -\frac{1}{2} & 0 & -\frac{1}{\sqrt{10}} & 0 & -\frac{1}{2 \sqrt{5}}
& 0 \\
\ket{0,0,2,\frac{1}{2}}& 0 & -\frac{1}{\sqrt{3}} & \frac{1}{\sqrt{5}} & 0 & \frac{1}{\sqrt{15}} & 0 & 0 & -\frac{1}{2 \sqrt{5}} & 0 & -\frac{1}{\sqrt{10}} & 0 & -\frac{1}{2}
\\
\ket{0,1,1,-\frac{1}{2}}& 0 & 0 & 0 & 0 & 0 & -i \sqrt{\frac{2}{5}} & \frac{i}{\sqrt{2}} & 0 & 0 & 0 & -\frac{i}{\sqrt{10}} & 0 \\
\ket{0,1,1,\frac{1}{2}}& 0 & 0 & -i \sqrt{\frac{2}{5}} & 0 & 0 & 0 & 0 & \frac{i}{\sqrt{10}} & 0 & 0 & 0 & -\frac{i}{\sqrt{2}} \\
\ket{0,2,0,-\frac{1}{2}}& -\frac{1}{\sqrt{3}} & 0 & 0 & -\frac{1}{\sqrt{15}} & 0 & \frac{1}{\sqrt{5}} & \frac{1}{2} & 0 & -\frac{1}{\sqrt{10}} & 0 & \frac{1}{2 \sqrt{5}}
& 0 \\
\ket{0,2,0,\frac{1}{2}}& 0 & -\frac{1}{\sqrt{3}} & -\frac{1}{\sqrt{5}} & 0 & \frac{1}{\sqrt{15}} & 0 & 0 & \frac{1}{2 \sqrt{5}} & 0 & -\frac{1}{\sqrt{10}} & 0 & \frac{1}{2}
\\
\ket{1,0,1-\frac{1}{2}}& 0 & 0 & \frac{i}{\sqrt{10}} & 0 & i \sqrt{\frac{3}{10}} & 0 & 0 & i \sqrt{\frac{2}{5}} & 0 & \frac{i}{\sqrt{5}} & 0 & 0 \\
\ket{1,0,1,\frac{1}{2}}& 0 & 0 & 0 & -i \sqrt{\frac{3}{10}} & 0 & -\frac{i}{\sqrt{10}} & 0 & 0 & \frac{i}{\sqrt{5}} & 0 & i \sqrt{\frac{2}{5}} & 0 \\
\ket{1,1,0,-\frac{1}{2}}& 0 & 0 & \frac{1}{\sqrt{10}} & 0 & -\sqrt{\frac{3}{10}} & 0 & 0 & \sqrt{\frac{2}{5}} & 0 & -\frac{1}{\sqrt{5}} & 0 & 0 \\
\ket{1,1,0,\frac{1}{2}}& 0 & 0 & 0 & -\sqrt{\frac{3}{10}} & 0 & \frac{1}{\sqrt{10}} & 0 & 0 & \frac{1}{\sqrt{5}} & 0 & -\sqrt{\frac{2}{5}} & 0 \\
\ket{2,0,0,-\frac{1}{2}}& -\frac{1}{\sqrt{3}} & 0 & 0 & \frac{2}{\sqrt{15}} & 0 & 0 & 0 & 0 & \sqrt{\frac{2}{5}} & 0 & 0 & 0 \\
\ket{2,0,0,\frac{1}{2}}& 0 & -\frac{1}{\sqrt{3}} & 0 & 0 & -\frac{2}{\sqrt{15}} & 0 & 0 & 0 & 0 & \sqrt{\frac{2}{5}} & 0 & 0 \\
\noalign{\smallskip}\hline
\end{array}$
\end{tabular}
\end{table*}


\begin{table}
\caption{The shell model orbitals of the original spin-orbit like shells and of the proxy-SU(3) shells. The magic number $14$ is proposed as a sub-shell closure in Ref. \cite{Sorlin}. The symmetry of each proxy-SU(3) shell is U($\Omega$) with $\Omega={({N}+1)({N}+2)\over 2}$. The orbitals being replaced are denoted by bold letters. Adapted from Ref. \cite{EPJASM}.}\label{shells}
\begin{tabular}{ccc}
\noalign{\smallskip}\hline\noalign{\smallskip}
$U(\Omega)$   &                    &                \\
spin-orbit    &                    &                \\  
3D-HO         &                    &                \\
magic numbers & original orbitals & proxy orbitals   \\ 
 \noalign{\smallskip}\hline\noalign{\smallskip}
 
U(3)      & $1p^{1/2}_{\pm 1/2}$ & $1p^{1/2}_{\pm 1/2}$ \smallskip\\
6-14      &  $\bf 1d^{5/2}_{\pm 1/2,\pm 3/2}$&  $\bf 1p^{3/2}_{\pm 1/2,\pm 3/2}$   \smallskip\\
2-8       & $\bf 1d^{5/2}_{\pm 5/2}$ & - \bigskip\\
 
U(6)      & $2s^{1/2}_{\pm 1/2}$ & $2s^{1/2}_{\pm 1/2}$  \smallskip\\
14-28     & $1d^{3/2}_{\pm 1/2,\pm 3/2}$ & $1d^{3/2}_{\pm 1/2,\pm 3/2}$  \smallskip\\
8-20      & $\bf 1f^{7/2}_{\pm 1/2,\pm 3/2,\pm 5/2}$ & $\bf 1d^{5/2}_{\pm 1/2,\pm 3/2,\pm 5/2}$ \smallskip\\
          & $\bf 1f^{7/2}_{\pm 7/2}$ & - \bigskip\\

U(10)     & $2p^{1/2}_{\pm 1/2}$ & $2p^{1/2}_{\pm 1/2}$  \smallskip\\
28-50     & $2p^{3/2}_{\pm 1/2,\pm 3/2}$ &  $2p^{3/2}_{\pm 1/2,\pm 3/2}$  \smallskip\\
20-40     & $1f^{5/2}_{\pm 5/2, \pm3/2,\pm 1/2}$ & $1f^{5/2}_{\pm 5/2, \pm3/2,\pm 1/2}$  \smallskip\\
          & $\bf 1g^{9/2}_{\pm 1/2,..., \pm 7/2}$ & $\bf 1f^{7/2}_{\pm 1/2,..., \pm 7/2}$  \smallskip\\
          & $\bf 1g^{9/2}_{\pm 9/2}$ & - \bigskip\\

U(15)     & $3s^{1/2}_{\pm 1/2}$ &  $3s^{1/2}_{\pm 1/2}$ \smallskip\\
50-82     & $2d^{3/2}_{\pm 1/2,\pm 3/2}$ & $2d^{3/2}_{\pm 1/2,\pm 3/2}$  \smallskip\\
40-70     & $2d^{5/2}_{\pm 1/2,...,\pm 5/2}$ &  $2d^{5/2}_{\pm 1/2,...,\pm 5/2}$  \smallskip\\
          & $1g^{7/2}_{\pm 1/2,...,\pm 7/2}$ & $1g^{7/2}_{\pm 1/2,...,\pm 7/2}$ \smallskip\\
          & $\bf 1h^{11/2}_{\pm 1/2,...,\pm 9/2}$ & $\bf 1g^{9/2}_{\pm 1/2,...,\pm 9/2}$  \smallskip\\
          & $\bf 1h^{11/2}_{\pm 11/2}$ & -  \bigskip\\
 
U(21)     & $3p^{1/2}_{\pm 1/2}$ & $3p^{1/2}_{\pm 1/2}$  \smallskip\\
82-126    & $3p^{3/2}_{\pm 1/2,\pm 3/2}$ & $3p^{3/2}_{\pm 1/2,\pm 3/2}$ \smallskip\\
70-112    & $2f^{5/2}_{\pm 1/2,...,\pm 5/2}$ & $2f^{5/2}_{\pm 1/2,...,\pm 5/2}$  \smallskip\\
          & $2f^{7/2}_{\pm 1/2,...,\pm 7/2}$ & $2f^{7/2}_{\pm 1/2,...,\pm 7/2}$  \smallskip\\
          & $1h^{9/2}_{\pm 1/2,...,\pm 9/2}$ & $1h^{9/2}_{\pm 1/2,...,\pm 9/2}$ \smallskip\\
          & $\bf 1i^{13/2}_{\pm 1/2,...,\pm 11/2}$ & $\bf 1h^{11/2}_{\pm 1/2,...,\pm 11/2}$  \smallskip\\
          & $\bf 1i^{13/2}_{\pm 13/2}$ & -  \bigskip \\
 
U(28)     & $4s^{1/2}_{\pm 1/2}$ & $4s^{1/2}_{\pm 1/2}$  \smallskip\\
126-184   & $3d^{3/2}_{\pm 1/2,\pm 3/2}$ &  $3d^{3/2}_{\pm 1/2,\pm 3/2}$ \smallskip\\
112-168   & $3d^{5/2}_{\pm 1/2,...,\pm 5/2}$ & $3d^{5/2}_{\pm 1/2,...,\pm 5/2}$  \smallskip\\
          & $2g^{7/2}_{\pm 1/2,...,\pm 7/2}$ &  $2g^{7/2}_{\pm 1/2,...,\pm 7/2}$ \smallskip\\
          & $2g^{9/2}_{\pm 1/2,...,\pm 9/2}$ &  $2g^{9/2}_{\pm 1/2,...,\pm 9/2}$  \smallskip\\
          & $1i^{11/2}_{\pm 1/2,...,\pm 11/2}$ & $1i^{11/2}_{\pm 1/2,...,\pm 11/2}$  \smallskip\\
          & $\bf 1j^{15/2}_{\pm 1/2,...,\pm 13/2}$ & $\bf 1i^{13/2}_{\pm 1/2,...,\pm 13/2}$  \smallskip\\
          & $\bf 1j^{15/2}_{\pm 15/2}$ & - \bigskip\\

\noalign{\smallskip}\hline\noalign{\smallskip}
\end{tabular}
\end{table}

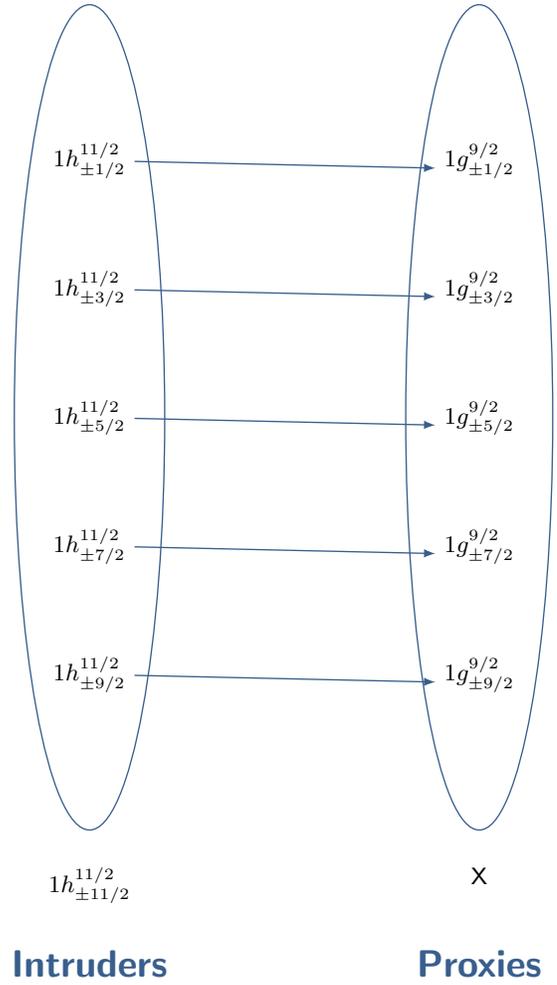
\begin{figure}
\begin{tikzpicture}[line width=0.5pt,>=latex]
\sffamily
\node (a1) {$1h^{11/2}_{\pm 1/2}$};
\node[below=of a1] (a2) {$1h^{11/2}_{\pm 3/2}$};
\node[below=of a2] (a3) {$1h^{11/2}_{\pm 5/2}$};
\node[below=of a3] (a4) {$1h^{11/2}_{\pm 7/2}$};
\node[below=of a4] (a5) {$1h^{11/2}_{\pm 9/2}$};

\node[right=4cm of a1] (b1) {$1g^{9/2}_{\pm 1/2}$};
\node[below= of b1] (b2) {$1g^{9/2}_{\pm 3/2}$};
\node[below=of b2] (b3) {$1g^{9/2}_{\pm 5/2}$};
\node[below=of b3] (b4) {$1g^{9/2}_{\pm 7/2}$};
\node[below=of b4] (b5) {$1g^{9/2}_{\pm 9/2}$};

\node[shape=ellipse,draw=myblue,minimum size=1cm,fit={(a1) (a5)}] {};
\node[shape=ellipse,draw=myblue,minimum size=1cm,fit={(b1) (b5)}] {};

\node[below=2.1cm of a5] {$1h^{11/2}_{\pm 11/2}$};
\node[below=2.1cm of b5] {X};

\node[below=3.2cm of a5,font=\color{myblue}\Large\bfseries] {Intruders};
\node[below=3.2cm of b5,font=\color{myblue}\Large\bfseries] {Proxies};

\draw[->,myblue] (a1) -- (b1.190);
\draw[->,myblue] (a2) -- (b2.190);
\draw[->,myblue] (a3) -- (b3.190);
\draw[->,myblue] (a4) -- (b4.190);
\draw[->,myblue] (a5) -- (b5.190);

\end{tikzpicture}
\caption{Unitary transformation of the intruder orbitals $1h^{11/2}_{m_j}$ (except for the $1h^{11/2}_{\pm 11/2}$) in the 50-82 shell onto the orbitals $1g^{9/2}_{m_j}$.
 Adapted from Ref. \cite{EPJASM}.}\label{proxies}
\end{figure}

\begin{table*}
\centering
\caption{Expansions of Nilsson orbitals $K[N n_z \Lambda]$ in the shell model basis $|N l j m_j \rangle$ for three different values of the deformation $\epsilon$. The Nilsson orbitals shown possess the highest total angular momentum $j$ in their shell. The existence of a leading shell model eigenvector is evident at all deformations.  Adapted from Ref. \cite{EPJPSM}. 
} \label{Hadi1}
\begin{tabular}{ r c r r r r r r  }
\hline\noalign{\smallskip}
${3\over 2}[541]$ & $\epsilon$ & $|N l j m_j \rangle$ & 
$\left| 5 1 {3\over 2} {3\over 2} \right\rangle$ & 
$\left| 5 3 {5\over 2} {3\over 2} \right\rangle$ & 
$\left| 5 3 {7\over 2} {3\over 2} \right\rangle$ & 
$\left| 5 5 {9\over 2} {3\over 2} \right\rangle$ & 
$\left| 5 5 {11\over 2} {3\over 2} \right\rangle$ \\

\noalign{\smallskip}\hline\noalign{\smallskip}
 & 0.05 & & 0.0025 & $-0.0015$ & 0.0641 & $-0.0122$ & 0.9979 \\
 & 0.22 & & 0.0371 & $-0.0286$ & 0.2565 & $-0.0640$ & 0.9633 \\
 & 0.30 & & 0.0601 & $-0.0506$ & 0.3287 & $-0.0922$ & 0.9366 \\

\noalign{\smallskip}\hline
\end{tabular}

\begin{tabular}{ r c  r r r r r r r }
\hline\noalign{\smallskip}
${3\over 2}[651]$ & $\epsilon$ & $|N l j m_j \rangle$ & 
$\left| 6 2 {3\over 2} {3\over 2} \right\rangle$ & 
$\left| 6 2 {5\over 2} {3\over 2} \right\rangle$ & 
$\left| 6 4 {7\over 2} {3\over 2} \right\rangle$ & 
$\left| 6 4 {9\over 2} {3\over 2} \right\rangle$ & 
$\left| 6 6 {11\over 2} {3\over 2} \right\rangle$ &
$\left| 6 6 {13\over 2} {3\over 2} \right\rangle$ \\

\noalign{\smallskip}\hline\noalign{\smallskip}

 & 0.05 & & $-0.0002$ & 0.0046 & $-0.0013$ & 0.0821 & $-0.0086$ & 0.9966  \\
 & 0.22 & & $-0.0100$ & 0.0711 & $-0.0278$ & 0.3240 & $-0.0469$ & 0.9418  \\
 & 0.30 & & $-0.0207$ & 0.1149 & $-0.0509$ & 0.4091 & $-0.0687$ & 0.9010  \\

\noalign{\smallskip}\hline
\end{tabular}

\end{table*}

\begin{table*}
\centering
\caption{Expansions of Nilsson orbitals $K[N n_z \Lambda]$ in the shell model basis $|N l j m_j \rangle$ for three different values of the deformation $\epsilon$. The Nilsson orbitals shown do not possess the highest total angular momentum $j$ in their shell. The existence of a leading shell model eigenvector is evident at small deformation, but this is not the case any more at higher deformations, at which several shell model eigenvectors make considerable contributions. Adapted from Ref. \cite{EPJPSM}.
}  \label{Hadi2}
\begin{tabular}{r c  r r r r r r  }
\hline\noalign{\smallskip}
${1\over 2}[431]$ & $\epsilon$ & $|N l j m_j \rangle$ & 
$\left| 4 0 {1\over 2} {1\over 2} \right\rangle$ & 
$\left| 4 2 {3\over 2} {1\over 2} \right\rangle$ & 
$\left| 4 2 {5\over 2} {1\over 2} \right\rangle$ & 
$\left| 4 4 {7\over 2} {1\over 2} \right\rangle$ & 
$\left| 4 4 {9\over 2} {1\over 2} \right\rangle$ \\

\noalign{\smallskip}\hline\noalign{\smallskip}
 & 0.05 & & $-0.0213$ & 0.1254 & $-0.0702$ & 0.9893 & 0.0127 \\
 & 0.22 & & $-0.2248$ & 0.4393 & $-0.2791$ & 0.8057 & 0.1717 \\
 & 0.30 & & $-0.2630$ & 0.5003 & $-0.2458$ & 0.7447 & 0.2559 \\

\noalign{\smallskip}\hline
\end{tabular}

\begin{tabular}{ r c r r r r r r r }
\hline\noalign{\smallskip}
${1\over 2}[541]$ & $\epsilon$ & $|N l j m_j \rangle$ & 
$\left| 5 1 {1\over 2} {1\over 2} \right\rangle$ & 
$\left| 5 1 {3\over 2} {1\over 2} \right\rangle$ & 
$\left| 5 3 {5\over 2} {1\over 2} \right\rangle$ & 
$\left| 5 3 {7\over 2} {1\over 2} \right\rangle$ & 
$\left| 5 5 {9\over 2} {1\over 2} \right\rangle$ &
$\left| 5 5 {11\over 2} {1\over 2} \right\rangle$ \\

\noalign{\smallskip}\hline\noalign{\smallskip}

 & 0.05 & & $-0.0200$ & 0.1770 & $-0.0295$ & 0.9780 & $-0.0446$ & $-0.0944$  \\
 & 0.22 & & $-0.2492$ & 0.4619 & $-0.3768$ & 0.5550 & $-0.4161$ & $-0.3185$  \\
 & 0.30 & & $-0.3121$ & 0.4331 & $-0.4829$ & 0.3430 & $-0.4789$ & $-0.3671$  \\

\noalign{\smallskip}\hline
\end{tabular}

\end{table*}

\section{Connecting the cartesian Elliott basis to the spherical shell model basis} 

In the Elliott model \cite{Elliott1,Elliott2,Elliott3,Elliott4,Harvey} the cartesian basis of the 3D isotropic HO is used, $[n_z n_x n_y m_s]$, in which the number of quanta along the $z$, $x$, $y$ directions and the 
$z$-projection of the spin appear. This basis can be transformed to the spherical basis $[n l m_l m_s]$ in $l$-$s$ coupling, labeled by the principal quantum number $n$, the orbital angular momentum $l$, its $z$-projection ($m_l$), and the $z$-projection of the spin ($m_s$), through a unitary transformation \cite{Davies,Chasman,Chacon} 
\begin{equation}
[n_z n_x n_y m_s] = R [n l m_l m_s], 
\end{equation}
the details of which can be found in Ref. \cite{EPJASM}. Using Clebsch-Gordan coefficients  the spherical basis can be rewritten as 
\begin{equation}
[n l m_l m_s] =  C [n l j m_j],  
\end{equation}
in which the total angular momentum $j$ and its $z$-projection appear. Combining these two transformations one obtains 
\begin{equation}
[n_z n_x n_y m_s] = R  C [n l j m_j],  
\end{equation}
i.e., the connection between the cartesian Elliott basis and the shell model basis in $j$-$j$ coupling. An example of this transformation is shown in Table \ref{N2}. Details of the calculations and tables for other shells can be found in Ref. \cite{EPJASM}. 

Using the above transformation one sees that the Nilsson 0[110] replacements made within the proxy-SU(3) scheme are ``translated'' into $| 0 1 1 0\rangle$ replacements within the spherical shell model basis. The resulting correspondence between  original shell model orbitals and proxy-SU(3) orbitals is summarized in Table \ref{shells}. This correspondence paves the way for taking advantage of the proxy-SU(3) symmetry in shell model calculations for heavy nuclei, in a way similar to that of the symmetry-adapted no-core shell model approach \cite{Launey1,Launey2} used in light nuclei. 

The correspondence between Nilsson pairs and shell model pairs has been corroborated by calculations \cite{EPJPSM} within the Nilsson model, in which the first justification of the proxy-SU(3) scheme has been found \cite{proxy1}. As one can see in Tables \ref{Hadi1} and \ref{Hadi2}, the correspondence used in proxy-SU(3) works only for the Nilsson orbitals which possess the highest total angular momentum $j$ within their shell, which are exactly the orbitals which are replaced within the proxy-SU(3) scheme. In further corroboration of this result, a unitary transformation connecting the orbitals being replaced within the proxy-SU(3) scheme has been found \cite{EPJASM} within the shell model basis and is depicted in Fig. 1. 

A by-product of the above transformation is that the 0[110] Nilsson pairs identified in Ref. \cite{Burcu} and used within the proxy-SU(3) scheme \cite{proxy1,proxy2,proxy3} are identical to the de Shalit--Goldhaber pairs \cite{deShalit} and the Federman--Pittel pairs \cite{FP1,FP2,FP3} within the spherical shell model basis, in which they are expressed as $| 0 1 1 0\rangle$ pairs. 

Within the proxy-SU(3) scheme the importance of the highest weight irreducible representations of SU(3) has been demonstrated \cite{EPJAHW} and used \cite{proxy2,proxy3} for the successful prediction of the prolate to oblate shape transition at $N=114$, the dominance of prolate over oblate shapes in the ground states of even-even nuclei, and the prediction of specific islands on the nuclear chart in which shape coexistence can appear \cite{EPJASC}. The compatibility of certain predictions of proxy-SU(3) and pseudo-SU(3) has been demonstrated in Refs. \cite{Cseh,EPJST}. These topics will be discussed in the talks by A. Martinou \cite{Martinou} and S. Sarantopoulou \cite{Sarantopoulou}.


\begin{thebibliography}{}

\bibitem{Kota}
V.K.B. Kota, \textit{SU(3) Symmetry in Atomic Nuclei} (Springer, Singapore, 2020)

\bibitem{Mayer1}
M. G. Mayer, Phys. Rev. \textbf{74}, 235 (1948)

\bibitem{Mayer2}
M. G. Mayer, Phys. Rev. \textbf{75}, 1969 (1949)

\bibitem{Jensen}
O. Haxel, J.H.D. Jensen, and H.E. Suess, Phys. Rev. \textbf{75}, 1766 (1949)

\bibitem{MJ}
M.G. Mayer and J.H.D. Jensen, {\it Elementary Theory of Nuclear Shell Structure} (Wiley, New York, 1955)  

\bibitem{Wybourne}
B. G. Wybourne, {\it Classical Groups for Physicists} (Wiley, New York, 1974)

\bibitem{Smirnov}
M. Moshinsky and Yu. F. Smirnov, {\it The Harmonic Oscillator in Modern Physics} (Harwood, Amsterdam, 1996) 

\bibitem{IacLie}
F. Iachello, {\it Lie Algebras and Applications} (Springer, Berlin, 2006)

\bibitem{BK}
D. Bonatsos and A. Klein, Ann. Phys. (NY) \textbf{169}, 61 (1986)

\bibitem{Bohr}
A. Bohr, Mat. Fys. Medd. K. Dan. Vidensk. Selsk. \textbf{26}, no. 14 (1952)

\bibitem{BM}
A. Bohr and B. R. Mottelson, {\it Nuclear Structure Vol. II: Nuclear Deformations} (Benjamin, New York, 1975)

\bibitem{Nilsson1}
S.G. Nilsson, Mat. Fys. Medd. K. Dan. Vidensk. Selsk.  \textbf{29}, no. 16 (1955)

\bibitem{NR}
S.G. Nilsson and I. Ragnarsson, {\it Shapes and Shells in Nuclear Structure} (Cambridge University Press, Cambridge, 1995)

\bibitem{Takahashi}
Y. Takahashi, Prog. Theor. Phys. \textbf{53}, 461 (1975)

\bibitem{Asherova}
R.M. Asherova, Yu.F. Smirnov, V.N. Tolstoy, and A.P. Shustov, Nucl. Phys. A \textbf{355}, 25 (1981)

\bibitem{RD}
G. Rosensteel and J. P. Draayer, J. Phys. A: Math. Gen. \textbf{22}, 1323 (1989)

\bibitem{ND}
W. Nazarewicz and J. Dobaczewski,  Phys. Rev. Lett. \textbf{68}, 154 (1992)

\bibitem{PVI}
W. Nazarewicz, J. Dobaczewski, and P. Van Isacker, AIP Conf. Proc. \textbf{259}, 30 (1992)

\bibitem{Lenis}
D. Bonatsos, C. Daskaloyannis, P. Kolokotronis, and D. Lenis, arXiv: hep-th/9411218   

\bibitem{Elliott1}
J. P. Elliott, Proc. Roy. Soc. London Ser. A  \textbf{245}, 128 (1958) 
  
\bibitem{Elliott2}
J. P. Elliott, Proc. Roy. Soc. London Ser. A  \textbf{245}, 562 (1958) 

\bibitem{Elliott3}
J. P. Elliott and M. Harvey, Proc. Roy. Soc. London Ser. A  \textbf{272}, 557 (1963)

\bibitem{Elliott4}
J. P. Elliott and C. E. Wildson, Proc. Roy. Soc. London Ser. A  \textbf{302}, 509 (1968)

\bibitem{Harvey}
M. Harvey, in Adv. Nucl. Phys. \textbf{1}, ed. M. Baranger and E. Vogt (Plenum, New York, 1968) p. 67 

\bibitem{Raychev25}
P.P. Raychev, Compt. Rend. Acad. Bulg. Sci. \textbf{25}, 1503 (1972)

\bibitem{Afanasev}
G.N. Afanas'ev, S.A. Abramov, and P.P. Raychev, Yad. Fiz. \textbf{16}, 53 (1972) [Sov. J. Nucl. Phys. \textbf{16}, 27 (1973)] 

\bibitem{Raychev16}
P.P. Raychev, Yad. Fiz. \textbf{16}, 1171 (1972) [Sov. J. Nucl. Phys. \textbf{16}, 643 (1973)]

\bibitem{RR27}
P.P. Raychev and R.P. Roussev, Yad. Fiz. \textbf{27}, 1501 (1978) [Sov. J. Nucl. Phys. \textbf{27}, 792 (1978)]
 
\bibitem{Minkov1}
N. Minkov, S.B. Drenska, P.P. Raychev, R.P. Roussev, and D. Bonatsos, Phys. Rev. C \textbf{55}, 2345 (1997)

\bibitem{Minkov2}
N. Minkov, S.B. Drenska, P.P. Raychev, R.P. Roussev, and D. Bonatsos, Phys. Rev. C \textbf{60}, 034305 (1999)

\bibitem{Minkov3}
N. Minkov, S.B. Drenska, P.P. Raychev, R.P. Roussev, and D. Bonatsos, Phys. Rev. C \textbf{61}, 064301 (2000)

\bibitem{AR}
G.N. Afanas'ev and P.P. Raychev, Fiz. Elem. Chast. At. Yadra \textbf{3}, 436 (1972) [Sov. J. Nucl. Phys. \textbf{3}, 229 (1972)]

\bibitem{Adler}
K.T. Hecht and A. Adler, Nucl. Phys. A \textbf{137}, 129 (1969)

\bibitem{Shimizu} 
A. Arima, M. Harvey, and K. Shimizu, Phys. Lett. B \textbf{30}, 517 (1969)

\bibitem{pseudo1}
R. D. Ratna Raju, J. P. Draayer, and K. T. Hecht, Nucl. Phys. A \textbf{202}, 433 (1973)

\bibitem{pseudo2}
J. P. Draayer, K. J. Weeks, and K. T. Hecht, Nucl. Phys. A \textbf{381}, 1 (1982)

\bibitem{DW1}
J. P. Draayer and K. J. Weeks, Phys. Rev. Lett. \textbf{51}, 1422 (1983)

\bibitem{DW2}
J. P. Draayer and K. J. Weeks, Ann. Phys. (N.Y.) \textbf{156}, 41 (1984)

\bibitem{Harwood} 
J.P. Draayer, in \textit{Algebraic Approaches to Nuclear Structure}, ed. R.F. Casten (Harwood, Chur, 1993), p. 423

\bibitem{AnnArbor}
O. Casta\~nos, M. Moshinsky, and C. Quesne, {\it Group Theory and Special Symmetries in Nuclear Physics Ann Arbor, 1991}, ed. J. P. Draayer and J. J\"anecke (World Scientific, Singapope, 1992) p. 80 

\bibitem{Quesne}  
O. Casta\~nos, M. Moshinsky, and C. Quesne, Phys. Lett. B \textbf{277}, 238 (1992)

\bibitem{Hess}
O. Casta\~nos, V. Vel\'azquez A., P.O. Hess, and J.G. Hirsch, Phys. Lett. B \textbf{321}, 303 (1994)

\bibitem{Ginocchio1}
J.N. Ginocchio, Phys. Rev. Lett. \textbf{78}, 436 (1997)

\bibitem{Ginocchio2}
J.N. Ginocchio, J. Phys. G: Nucl. Part. Phys. \textbf{25}, 617 (1999)

\bibitem{Jolos}
D. Janssen, R.V. Jolos, and F. D\"onau, Nucl. Phys. A \textbf{224}, 93 (1974)

\bibitem{AI}
A. Arima and F. Iachello, Phys. Rev. Lett \textbf{35}, 1069 (1975)

\bibitem{IA}
F. Iachello and A. Arima, {\it The Interacting Boson Model} (Cambridge University Press, Cambridge, 1987)

\bibitem{IVI}
F. Iachello and P. Van Isacker, \textit{The Interacting Boson-Fermion Model} (Cambridge University Press, Cambridge, 1991)

\bibitem{FVI}
A. Frank and P. Van Isacker, \textit{Symmetry Methods in Molecules and Nuclei} (S y G editores, M\'exico D.F., 2005)

\bibitem{Rosensteel}
G. Rosensteel and D.J. Rowe, Ann. Phys. (NY) \textbf{126}, 343 (1980)

\bibitem{RW}
D. J. Rowe and J. L. Wood, \textit{Fundamentals of Nuclear Models: Foundational Models} (World Scientific, Singapore, 2010)

\bibitem{IVBM1}
A. Georgieva, P. Raychev, and R. Roussev, J. Phys. G: Nucl. Phys. \textbf{8}, 1377 (1982)

\bibitem{IVBM2}
A. Georgieva, P. Raychev, and R. Roussev, J. Phys. G: Nucl. Phys. \textbf{9}, 521 (1983)

\bibitem{FDSM} 
C.-L. Wu, D. H. Feng, X.-G. Chen, J.-Q. Chen, and M. W. Guidry, Phys. Rev. C \textbf{36}, 1157 (1987)

\bibitem{proxy1}
D. Bonatsos, I. E. Assimakis, N. Minkov, A. Martinou, R. B. Cakirli, R. F. Casten, and K. Blaum, Phys. Rev. C \textbf{95}, 064325 (2017)

\bibitem{proxy2}
D. Bonatsos, I. E. Assimakis, N. Minkov, A. Martinou, S. Sarantopoulou, R. B. Cakirli, R. F. Casten, and K. Blaum, Phys. Rev. C \textbf{95}, 064326 (2017)

\bibitem{proxy3}
D. Bonatsos, Eur. Phys. J. A \textbf{53}, 148 (2017)

\bibitem{deShalit}
A. de Shalit and M. Goldhaber, Phys. Rev. \textbf{92}, 1211 (1953)

\bibitem{Talmi62}
I. Talmi, Rev. Mod. Phys. \textbf{34}, 704 (1962)

\bibitem{Talmi71}
I. Talmi, Nucl. Phys. A \textbf{172}, 1 (1971)

\bibitem{Talmi73}
I. Talmi, Riv. Nuovo Cim. \textbf{3}, 85 (1973)

\bibitem{Talmi93}
I. Talmi, \textit{Simple Models of Complex Nuclei} (Harwood, Chur, 1993)


\bibitem{FP1}
P. Federman and S. Pittel, Phys. Lett. B \textbf{69}, 385 (1977) 

\bibitem{FP2}
P. Federman and S. Pittel, Phys. Lett. B \textbf{77}, 29 (1978)

\bibitem{FP3}
P. Federman and S. Pittel, Phys. Rev. C \textbf{20}, 820 (1979)

\bibitem{CastenPRL}
R.F. Casten, Phys. Rev. Lett. \textbf{54}, 1991 (1985)

\bibitem{Casten}
R.F. Casten, Nucl. Phys. A \textbf{443}, 1 (1985)

\bibitem{Haustein}
R.F. Casten, D.S. Brenner, and P.E. Haustein,  Phys. Rev. Lett. \textbf{58}, 658 (1987)

\bibitem{Castenbook}
R. F. Casten, \textit{Nuclear Structure from a Simple Perspective} (Oxford University Press, Oxford, 2000)

\bibitem{Zuker1}
A. P. Zuker, J. Retamosa, A. Poves, and E. Caurier, Phys. Rev. C \textbf{52}, R1741 (1995)

\bibitem{Zuker2}
A. P. Zuker, A. Poves, F. Nowacki, and S. M. Lenzi, Phys. Rev. C \textbf{92}, 024320 (2015) 

\bibitem{Kaneko}
K. Kaneko, N. Shimizu, T. Mizusaki, and Y. Sun, Phys. Rev. \textbf{103}, L021301 (2021)

\bibitem{Cakirli94}
R.B. Cakirli, D.S. Brenner, R.F. Casten, and E.A. Millman, Phys. Rev. Lett. \textbf{94}, 092501 (2005); erratum Phys. Rev. Lett. \textbf{95}, 119903 (2005)

\bibitem{Cakirli96}
R.B. Cakirli and R.F. Casten, Phys. Rev. Lett. \textbf{96}, 132501 (2006)

\bibitem{Brenner}
D.S. Brenner, R.B. Cakirli and R.F. Casten, Phys. Rev. C \textbf{73}, 034315 (2006)

\bibitem{Stoitsov}
M. Stoitsov, R.B. Cakirli, R.F. Casten, W. Nazarewicz, and W. Satula, Phys. Rev. Lett. \textbf{98}, 132502 (2007)

\bibitem{Cakirli102}
R.B. Cakirli, R.F. Casten, R. Winkler, K. Blaum, and M. Kowalska, Phys. Rev. Lett. \textbf{102}, 082501 (2009)

\bibitem{Burcu}
R. B. Cakirli, K. Blaum and R. F. Casten, Phys. Rev. C \textbf{82}, 061304 (2010)
 
\bibitem{Sofia}
D. Bonatsos, S. Karampagia, R.B. Cakirli, R.F. Casten, K. Blaum,  and L. Amon Susam, Phys. Rev. C \textbf{88}, 054309 (2013)

\bibitem{Davies}
K.T.R. Davies and S.J. Krieger, Can. J. Phys. \textbf{69}, 62 (1991)

\bibitem{Chasman}
R.R. Chasman and S. Wahlborn, Nucl. Phys. A \textbf{90}, 401 (1967) 

\bibitem{Chacon}
E. Chac\'on and M. de Llano, Rev. Mex. F\'is. \textbf{12} 57 (1963)

\bibitem{EPJASM}
A. Martinou, D. Bonatsos, N. Minkov, I.E. Assimakis, S.K. Peroulis, S. Sarantopoulou, and J. Cseh, Eur. Phys. J. A \textbf{56}, 239 (2020)

\bibitem{Sorlin}
O. Sorlin and M.-G. Porquet, Prog.  Part. Nucl. Phys. \textbf{61}, 602 (2008)

\bibitem{Launey1}
K.D. Launey, J.P. Draayer, T. Dytrych, G.-H. Sun, and S.-H. Dong, Int. J. Mod. Phys. E \textbf{24}, 1530005 (2015)
 
\bibitem{Launey2}
K.D. Launey, T. Dytrych, and J.P. Draayer, Prog. Part. Nucl. Phys. \textbf{95}, 044312 (2017)

\bibitem{EPJPSM}
D. Bonatsos, H. Sobhani, and H. Hassanabadi, Eur. Phys. J. Plus \textbf{135}, 710 (2020)

\bibitem{EPJAHW}
A. Martinou, D. Bonatsos, S. Sarantopoulou, I.E. Assimakis, S.K. Peroulis, and N. Minkov, Eur. Phys. J. A \textbf{57}, 83 (2021)

\bibitem{EPJASC}
A. Martinou, D. Bonatsos, T.J. Mertzimekis, K. Karakatsanis, I.E. Assimakis, S.K. Peroulis, S. Sarantopoulou, and N. Minkov, Eur. Phys. J. A  \textbf{57}, 84 (2021)

\bibitem{Cseh} J. Cseh, Phys. Rev. C \textbf{101}, 054306 (2020)

\bibitem{EPJST}
D. Bonatsos, A. Martinou, S. Sarantopoulou, I.E. Assimakis, S. Peroulis, and N. Minkov, Eur. Phys. J. ST  \textbf{229}, 2367 (2020)

\bibitem{Martinou}
A. Martinou \textit{et al.}, these proceedings 

\bibitem{Sarantopoulou}
S. Sarantopoulou \textit{et al.}, these proceedings 

\end{thebibliography}
\end{document}